# A Survey on Hardware Implementations of Visual Object Trackers

Al-Hussein A. El-Shafie and S. E. D. Habib

*Abstract*— **Visual object tracking is an active topic in the computer vision domain with applications extending over numerous fields. The main sub-tasks required to build an object tracker (e.g. object detection, feature extraction and object tracking) are computation-intensive. In addition, real-time operation of the tracker is indispensable for almost all of its applications. Therefore, complete hardware or hardware/software co-design approaches are pursued for better tracker implementations. This paper presents a literature survey of the hardware implementations of object trackers over the last two decades. Although several tracking surveys exist in literature, a survey addressing the hardware implementations of the different trackers is missing. We believe this survey would fill the gap and complete the picture with the existing surveys of how to design an efficient tracker and point out the future directions researchers can follow in this field. We highlight the lack of hardware implementations for state-of-the-art tracking algorithms as well as for enhanced classical algorithms. We also stress the need for measuring the tracking performance of the hardware-based trackers. Additionally, enough details of the hardware-based trackers need to be provided to allow reasonable comparison between the different implementations.**

*Index Terms*—**object tracking, hardware implementation, computer vision, video processing, FPGA, SoC, VLSI, hardware/software co-design**

## I. INTRODUCTION

Visual object tracking is the process of estimating the location of one or more objects in the video frames. Tracking has wide application domains like surveillance systems, intelligent robotics, unmanned vehicles and virtual reality. The real-world challenges such as occlusion, noise, changing appearance, cluttered background, and illumination variations still keep the tracking problem an open and active research topic in the computer vision domain [1]. In addition, the dramatic increase in computational power over the last two decades has opened the door for the creation of new algorithms and applications.

In general, the main steps required to build a tracking system are object detection, appearance modeling and tracking [2]. The object detection step may be avoided and a user-specified target is employed instead. For each of these tracking steps, numerous algorithms exist in literature and are demonstrated in real systems aiming to improve the overall tracking performance. Desktop computers can be employed to execute such algorithms and can even achieve real-time operation. However, high power consumption, size and mobility of the desktop computers pretty much forbid their deployment in practical tracking systems. They are used mainly for profiling and validation. Hence, embedded platforms are the typical choice for implementing the tracking algorithms. The embedded implementations can be intuitively classified as complete software (SW), hardware/software (HW/SW) co-design and complete hardware (HW) approaches. Several SW implementations exist in literature targeting Digital Signal Processors (DSP) [3], [4], Graphics Processing Units (GPU) [5], [6] and embedded Central Processing Units (CPU) [7], [8]. This category features high-degree of flexibility and fast development time. Adding a HW accelerator to an embedded processor unit to handle time-consuming tasks is the second category: HW/SW co-design [9]-[12]. The execution time of the algorithm is improved accordingly. The last category is to implement the tracking algorithms completely in HW [13]-[16]. This category would give the best performance, area and power consumption compared to the other categories. Because of the complexity of the tracking algorithms and the speed and power constraints of most tracking applications, we believe that complete HW and HW/SW co-design approaches are well-suited for the tracker implementation.

Several reviews of the different algorithms used in the object trackers are published. Since the tracking research field continues to be active since long period, one can expect such large number of surveys and even wait for at least a new survey every five years. However, despite the importance of the HW implementations in the tracking applications, a survey of the HW implementations of the different tracking algorithms does not exist yet in literature. Such survey, together with the existing algorithmic surveys of object tracking, would give a better picture of the current status of the tracking systems. Hence, we present in this paper a review of the published HW implementations of the different trackers over the last two decades to fill this gap.

Our paper objectives are three-fold:
1) To collect the literature on the HW implementations of object trackers. To the best of our knowledge, no review

A-H. A. El-Shafie is with the Faculty of Engineering, Cairo University, Giza, 12613, Egypt (e-mail: elshafie_a@yahoo.com).
S. E. D. Habib is with the Faculty of Engineering, Cairo University, Giza, 12613, Egypt (e-mail: seraged@ieee.org).

paper was previously published on this topic.
2) To propose new directions for HW implementations of object trackers based on recent progress on object tracking algorithms or based on new insights through classical algorithms.
3) To pave the way to a systematic approach for HW implementations of object trackers. Such approach should include the measurement of tracking performance of these trackers. Additionally, the reported HW details should suffice for direct comparisons across different platforms.

This paper is organized as follows: Section II describes a typical tracking system, the existing tacking surveys and our survey terms. The review of the different tracking algorithms based on our classification is presented in Section III with a recommendation of the future directions in each category. Section IV presents a general discussion about the HW implementation of the trackers, and finally, Section V concludes our work.

## II. OVERVIEW OF TRACKING SYSTEMS

A typical tracking system is shown in Fig. 1. A visual sensor or camera usually outputs the pixel values serially in a raster scan pattern. A pre-processing may be needed for serial-to-parallel conversion, color space conversion and noise removal. Then, it is required to detect and determine the objects to be tracked in the image sequences. This can be done automatically by object detection algorithms like background subtraction or temporal frame differencing. These techniques aim at segmenting foreground or moving objects from the background. They can be applied on the first frames only as an object detection step and then tracking algorithms work on the selected objects, or they can be applied on every frame and the tracking system output would be the detection step output directly, e.g. motion-detection based tracking. On the other hand, the user can manually choose the object location in the image with bounding boxes or ellipses. This would be the input for the tracking algorithm to work in the successive frames.

Once the target is determined, features can then be extracted from this target location for further tracking and matching steps. Features are also extracted in the next frames from the candidate locations obtained by the tracking algorithm in order to be compared with the target features. In general, features can be grouped into three main classes [17], low-level features like color and gradient, mid-level features like edges and corners and high-level features like objects.

Finally, the tacking algorithm determines the trajectory of the moving object for further control and higher-level actions. Ali et al. [21] classified the different tracking algorithms into classical and contemporary approaches. Classical approaches include mean-shift, filtering, correlation-based template matching and motion-detection based tracking. While, contemporary approaches include tracking by detection, particle swarm optimization and sparse representation.

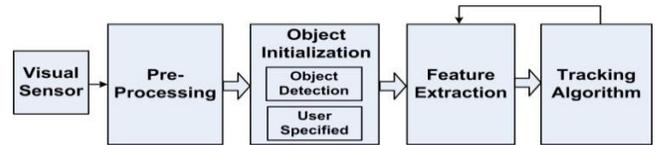
Fig. 1. A typical tracking system

As is expected there is no single tracking algorithm that is superior over all others in all aspects. There would be always trade-offs. Hence, the comparison between the tracking systems is challenging. Numerous surveys on tracking systems have been published in literature. Some review and compare general trackers [1], [18]-[21]. Others focus on specific applications like surveillance systems [22], [23], and pedestrian tracking [24], [25]. A few cover a specific stage or algorithm in the tracking system like object detection [26], [27], appearance models [2], [28], particle filters [29], [30], mean-shift [31] and optical flow [32]-[34].

Our survey reviews the HW implementation of the tracking systems. We focus on the published papers which present a tracking system with all or part of the system implemented in HW (e.g. HW and HW/SW co-design approaches). This would highlight the bottleneck in the different tracking algorithms based on the published work and point out how the selection of the object detection approach, the appearance model and the tracking algorithm would affect the system speed, area and memory size. For instance, based on the selected algorithm, the tracking steps can start directly when the camera outputs the frame pixels, and hence, save a lot in memory size. Otherwise, one or more frame or line buffers would be needed instead.

## III. REVIEW OF THE HARDWARE IMPLEMENTATIONS OF THE TRACKING SYSTEMS

We classify the published HW-based trackers, according to the tracking algorithm, in six categories: mean-shift, filtering techniques, feature matching, optical flow, template matching and bio-inspired techniques. In each category, we give a short overview of the tracking algorithm and its features. We list the published papers and highlight the visual features and the object detection approach employed in addition to the computation bottleneck in the whole system. The achieved frame rate, frame size, clock frequency and area would give insights about the strength of each implementation.

In general, we do not target a one-to-one comparison between the published HW-based trackers because most of the surveyed papers did not report the tracker performance. A unique performance measurement is indispensable for establishing a fair comparison between the HW-based trackers. In addition, for most of these papers, there is a lack of implementation details, like object size, Region Of Interest (ROI) size and details of the HW/SW partitioning. Instead, we focus in our survey on describing the HW architectures of the HW-based trackers and show how each paper attempts to overcome the performance bottleneck in the system.

## A. Mean-Shift based tracker

Mean-shift algorithm is a non-parametric estimation of a gradient density function. It is an iterative process that shifts a data point to the average of its neighborhood. Although mean-shift was not initially intended to be used as a tracking algorithm, it is quite effective in this role [35]. Comaniciu et al. [36] started the effort of using mean-shift technique in tracking non-rigid objects. By assuming that the target model has a density function, the problem is then to find the discrete location of the target candidate whose associated density function is the most similar to the target density function. This can be obtained by maximizing a similarity measure between the target density function and the candidate density function. Using Bhattacharyya coefficient as the similarity measure, Comaniciu et al. [36] shows that the Bhattacharyya coefficient depends on the density estimate computed at the candidate location in the current frame, and hence, maximization can be obtained by mean-shift iterations. Equation (1) describes the calculation of the mean-shift vector as given by [36].

$$m_{j+1} = \frac{\sum_{i=1}^{n} p_i \, w_i \, g\left(\left\|\frac{m_j - p_i}{h}\right\|^2\right)}{\sum_{i=1}^{n} w_i \, g\left(\left\|\frac{m_j - p_i}{h}\right\|^2\right)}, \quad j = 1, 2, .. \quad (1)$$

$$w_i = \sum_{u=1}^{l} \delta[b(p_i) - u] \sqrt{\frac{q_u}{p_u(m_j)}} \quad (2)$$

Where $m_{j+1}$ is the computed mean at iteration $j+1$ with kernel $G$ of window radius (bandwidth) $h$ and $\{p_i, w_i\}\, i=1..n$ are the sampled data points and their weights. $q_u$ and $p_u$ are density functions (e.g. color histogram of $l$-bins) of the target and candidate locations respectively. $\delta$ is the Dirac delta function. Fig. 2 illustrates the main steps in a mean-shift based tracker.

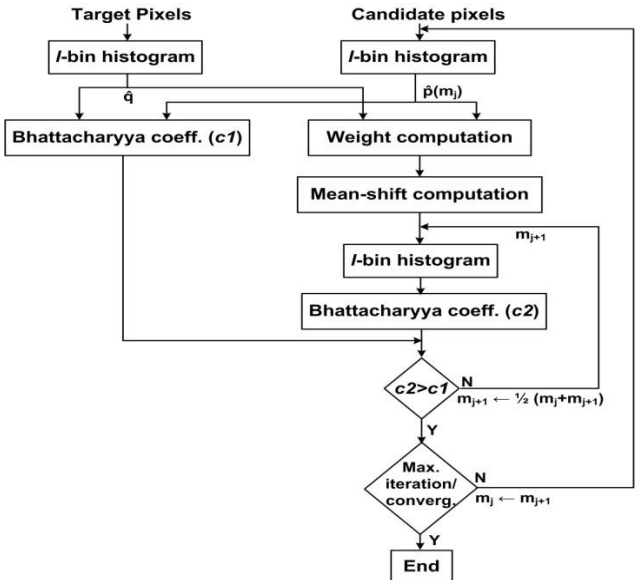

Fig. 2. Main steps in a mean-shift based tracker.

To avoid a potential issue in case of varying scale objects, different $h$ values can be tested and the value which achieves a better maximization is selected. On the other hand, color histogram is typically used as the density function of the target model and candidates. As color distribution can change over time, Bradski [35] proposes a Continuously Adaptive Mean Shift (CAMSHIFT) algorithm. The estimated size and location of the tracked object in the current frame are used to set the size and the location of the search window in the next frame. The center of the search window can be obtained by calculating the zeroth and the first moment features with the previous search window as described in equations (3), (4) and (5) for discrete 2D image probability distribution. This process is repeated till convergence is achieved. In addition, the size and the orientation of the search window can be obtained by calculating the second moment feature.

$$M_{00} = \sum_x \sum_y I(x, y) \quad (3)$$

$$M_{10} = \sum_x \sum_y x * I(x, y); \quad M_{01} = \sum_x \sum_y y * I(x, y) \quad (4)$$

$$x_c = \frac{M_{10}}{M_{00}}; \quad y_c = \frac{M_{01}}{M_{00}} \quad (5)$$

Where $I(x,y)$ is the pixel probability at position $(x,y)$, $M_{00}$ is the zeroth moment feature, $(M_{10}, M_{01})$ are the first moment features and $(x_c, y_c)$ is the new search window center.

Table I lists the HW-based mean-shift trackers published in literature and indicates whether they support Multi-object (MO) tracking and Automatic Detection (AD) for the objects or not. For the mean-shift implementations, Ali et al. [9] designed and integrated a co-processor with a MicroBlaze CPU to implement a color-histogram based mean shift algorithm. The co-processor handles most of the steps: image cropping and decimation step (from image resolution of 360x288 to 64x64), kernel computation, histogram, mean-shift displacement and Bhattacharyya coefficient calculations. The CPU was just employed for target initialization and control stuff. The authors adopted a histogram of 20 bins and a mean-shift with 20 iterations with scale adaptation. Taking the benefits of the pipelined divider and square root units, existing on the FPGA for the calculation of the mean-shift vector and Bhattacharyya coefficient, tremendously reduce the computation time. The proposed tracker can track a single target with maximum size of 64x64 at 290 Frame Per Second (fps) using a Spartan-3 FPGA. Multiple target tracking is also supported by replicating the mean-shift operation. While full SW implementation on a 50MHz MicroBlaze CPU achieves 10 fps for one target of 32x32 maximum size as presented by the same authors in [37], this HW/SW co-design approach achieves up to 833 fps for 32x32 single target tracking. Enabling scale adaptation and mutli-target tracking would reduce the performance as expected. Pandey et al. [10] presented pretty much a similar implementation but with no image cropping and with the Bhattacharya coefficient is computed in SW. They reported single-target tracking of maximum size of 160x80 at 60 fps using a Spartan-6 FPGA (the equivalent SW implementation achieves 5 fps).

By using the Logarithmic Number System (LNS), Pandey et al. [11] were able to avoid the complex operations, division and square root in the mean-shift weight and displacement computations. The weight computation in (2) is converted to four binary logarithmic units and one antilogarithmic unit in addition to simple addition operations as shown in Fig. 3. The displacement computation in (1) is converted to four binary logarithmic units and two antilogarithmic units in addition to addition and multiplication operations. This tracker is based on HW/SW co-design as well. An embedded PowerPC CPU was employed to control the frame capturing process and initialize the target location. The authors adopted a histogram of 4096-bin and three block RAMs of size 4096x32 to store the histogram of the target and the candidate in addition to the weight values. The proposed design achieves 60 fps for maximum object size of 64x64 on a Virtex-5 FPGA.

Norouznezhad et al. [12] proposed to use local oriented energy features instead of color features in order to increase the robustness of the tracker. Gabor filters across multiple channels can be used to extract the local oriented energy features. Hence, the authors proposed to convolve the target and candidate regions with 7x7 complex Gabor filters across 12 channels. To achieve that, six line buffers and two 7x7 filters (real and imaginary) with programmable coefficients were employed. The coefficients of the corresponding channel filter are loaded into the convolution unit which operates at much higher frequency than that of the incoming pixel stream, and hence, convolution across 12 channels can be obtained while using two filters only and saving area as shown in Fig. 4. Histograms of 12-bins are then calculated as the density function of the target and candidate for the mean-shift algorithm. The authors showed that the proposed system outperforms the color-based mean-shift implementation especially in case of a target with very similar color as the background. The Gabor filtering, local oriented energy feature extraction and feature histogram calculation were implemented in HW while the mean-shift algorithm was executed on a MicroBlaze CPU running at 125MHz. Using a Virtex-5 FPGA, the authors reported 30 fps while adopting eight mean-shift iterations.

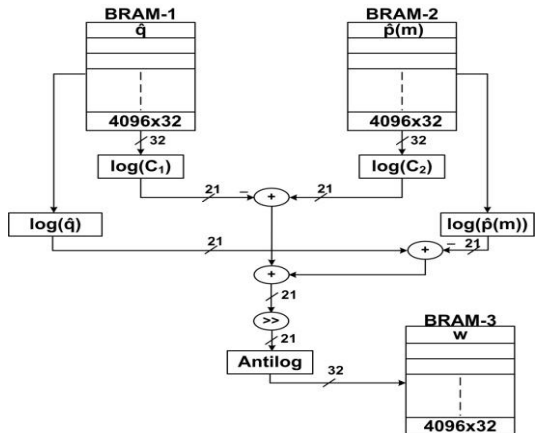

Fig. 3. Architecture of computing mean-shift weights using LNS [11]

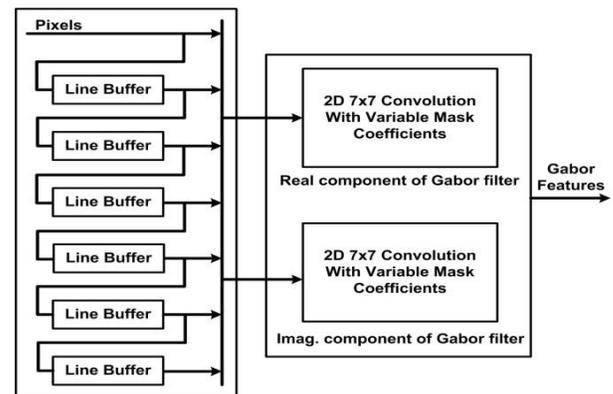

Fig. 4. Multi-channel complex Gabor filtering block [12]

TABLE I
MEAN-SHIFT BASED TRACKER WITH HW IMPLEMENTATION

| Year/ Tracker | Implementation | Visual feature | Image size | Maximum object size | Frame rate | HW Clock | MO | AD | Algorithm |
|---|---|---|---|---|---|---|---|---|---|
| 2010 [9] | HW: Spartan-3e SW: MicroBlaze | Color histogram | 360x288 | 64x64 | 290 fps | 50 MHz | ✕ | ✕ | Mean shift |
| 2016 [10] | HW: Spartan-6 SW: MicroBlaze | Color histogram | 1280x780 | 160x80 | 60 fps | 74.25 MHz | ✕ | ✕ | Mean shift |
| 2015 [11] | HW: Virtex-5 SW: PowerPC | Color histogram | 640x480 | 64x64 | 60 fps | NR* | ✕ | ✕ | Mean shift |
| 2010 [12] | HW: Virtex-5 SW: MicroBlaze | Local-oriented energy feature histogram | 640x480 | NR* | 30 fps | 296 MHz | ✕ | ✕ | Mean shift |
| 2008 [39] | HW: Virtex-4 SW: DSP[1] | Color histogram | 352x288 | 96x96 | 38 fps | 105 MHz | ✓ | ✓ | Camshift |
| 2010 [13] | HW: Cyclone | Color histogram | 720x576 | NR* | 25 fps | 100 MHz | ✕ | ✕ | Camshift |
| 2011 [38] | HW: Virtex-4 & Spartan-3 SW: PC, Core2 Q9300 | Color histogram | 512x511 | NR* | 2000 fps | 151.2 MHz | ✕ | ✕ | Camshift |

[1] hypothetical DSP
*NR: Not Reported in the original paper

On the other hand, for CAMSHIFT implementations, Lu et al. [13] proposed a CAMSHIFT-based tracking system with RGB to HSV color space conversion, Hue component histogram computation and moment calculation units. The authors adopted a pipeline structure for the RGB to HSV conversion on the pixel level and a pipeline structure for the Hue histogram computation on the image field level. The proposed structure would save the required memory storage. In addition, the multipliers and divisions are time-multiplexed between the different algorithm steps in order to save the HW resources. The zeroth and first moments are calculated and the center of the new search window is obtained. The authors reported a 25 fps performance using a Cyclone FPGA.

High frame rate tracking of 2000 fps was reported by Ishii et al. [38]. They noticed redundant multiplications when calculating the back-projection image and the moment features in (3), (4). Instead of calculating the back-projection image and then multiplying and summing over x and y coordinates, the authors proposed to calculate first the moment features for each bin of the binary image histogram, then sum them with the bin weights, and hence, same moment feature computation can be obtained with less complexity. The authors designed a system called IDP express that consists of a camera head and two FPGAs (Virtex-4 and Spartan-3) with FIFOs in between for data transfer. In addition, this board is connected to a PC via PCI-e. The image is sent from the camera as four parallel pixels in a raster scan fashion. On the FPGA, four parallel HW units for Bayer to RGB to HSV color conversion are employed. Moreover, 16 parallel moment calculation units are employed to obtain the six moment types for each bin of the total 16 bins used in the system as shown in Fig. 5. The six moment types form the zeroth, first and second moment features and allow calculating the center, size and orientation of new search window. The remaining steps of the algorithm were done by SW on the PC.

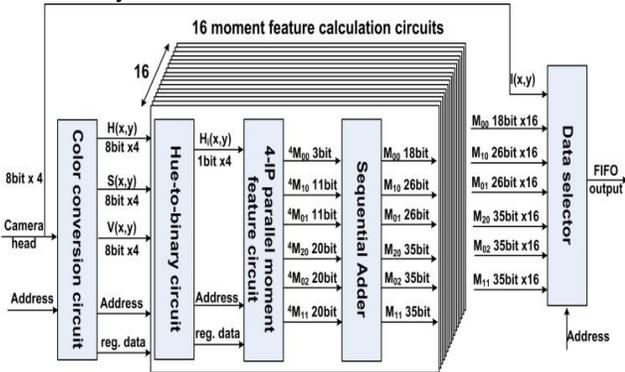

Fig. 5. Moment feature calculation [38]

Vijverberg et al. [39] presented a new technique to calculate the moments that is more suitable for HW implementation. The horizontal moments up to the second one are calculated for a number of consecutive pixels (four pixels in the paper) in a recursive way, then multiplication and summation over the coordinates are performed. The authors evaluated this technique as a co-processor implementation with a hypothetical DSP in a tracker case study based on mean-shift and CAMSHIFT approaches. The coprocessor handles Hue component to probability value LUT, histogram and moment feature computation. Implemented on a Virtex-4 FPGA, up to 16 objects of size 96x96 can be tracked with 38 fps and 42 fps for the CAMSHIFT and mean shift respectively. The authors adopted a 32-bin histogram and 10 mean-shift iterations.

To sum up the work on HW-based mean-shift trackers, most of the published trackers use color histograms to represent the target and candidate densities. Norouznezhad et al. [12] proposed to use the histogram of the local oriented energy features instead to overcome the color histogram limitations especially in case of similar object and background colors. It can be seen that the feature extraction and histogram calculation are typically implemented in HW because of the high processing rate required on the pixel level. The published trackers implemented the other steps such as the mean shift displacement, moments, search window update, and similarity computation in HW or SW depending on the design requirements. In general, the most complex operations in these steps are square root, division and multiplication operations. A proper implementation of these functions would significantly enhance the processing time. It should be noted also that the proposed designs in [12], [13], [38] process the candidate locations on the fly as the pixels arrive from the camera. This would not put a restriction on the maximum object size like the other designs do where the candidate pixels have to be fetched from the memory.

Some issues arise when using histogram as density representation and mean shift algorithm [21]. These issues include local maximum convergence and loss of spatial information. In addition, mean shift cannot handle occlusion, even if it is partial. The first two issues were handled in Shen et al. [40] and in Yang et al. [41] in which the kernel bandwidth is changed in an annealing fashion and a new similarity function in a joint spatial-features space is presented respectively. In addition, Jeyakar et al. [42] combined fragment-based approach with mean-shift to handle the partial occlusion scenario. We believe implementing these solutions in HW-based mean-shift trackers would enhance the overall system performance.

### B. Filtering techniques based tracker

In this category, statistical methods use the state space approach to model the object properties such as position, velocity and acceleration [18]. For a moving object in the scene, the location of the object can be defined by a sequence of states $X_k : k = 1, 2, \ldots$. The state transition and measurement equations can be given by (6) and (7) respectively.

$$X_k = f_k(X_{k-1}) + N_{1k} \qquad (6)$$

$$Z_k = h_k(X_k, N_{2k}) \qquad (7)$$

Where $Z_k$ is the measurement value at step $k$. $N_{1k}$ and $N_{2k}$ are the noise in the state and measurement equations. The objective of the tracking is to estimate the current state $X_k$ given all the measurements up to $k$ which is equivalent to obtain the conditional probability $P(X_k \mid Z_{1,\ldots,k})$. A recursive Bayesian filter can give an optimal solution by performing two steps, prediction and correction. The prediction step derives the prior probability density function (pdf) of the current state $P(X_k \mid Z_{1,\ldots,k-1})$. The correction step then employs the likelihood function $P(Z_k \mid X_k)$ to compute the posterior pdf $P(X_k \mid Z_{1,\ldots,k})$ [18]. If the functions $f_k$ and $h_k$ are linear and the initial state and noise have a Gaussian distribution, Kalman filter can give the optimal state estimation. In other words, Kalman filter is a recursive parametric estimation approach for discrete state space systems. Otherwise, if the functions $f_k$ and $h_k$ are nonlinear, Taylor series expansion can be used to linearize these functions [43], named extended Kalman filter. Still, the extended Kalman filter assumes Gaussian distribution of the state. However, the assumption of the Gaussian distribution of the states is not typically valid in case of general tracking systems, and thus, Kalman filter will give poor estimation for the states that do not follow Gaussian distribution [18]. Therefore, particle filtering [44] has been used instead to overcome this limitation. It uses Monte Carlo (MC) simulations to approximate the posterior density function by a set of random samples with associated weights as given by (8).

$$P(X_k \mid Z_{1:k}) \approx \sum_{i=1}^{N_s} W_k^i \, \delta(X_k - X_k^i) \quad (8)$$

Where $X_k^i, i = 1, \ldots, N_s$ are the random samples with the associated weights $W_k^i$ at time step $k$. As $N_s$ increases, the approximate density converges to the actual one. The principle of importance sampling is used to obtain the weights. If the samples can be drawn from another easy density, $q(X_k \mid Z_{1:k})$ which is called importance density, the weights can be calculated by (9).

$$W_k^i \propto \frac{P(X_k^i \mid Z_{1:k})}{q(X_k^i \mid Z_{1:k})} \quad (9)$$

Accordingly, obtaining samples (particles) from $q(X_k \mid Z_{1:k})$ and then weighting them by (9) is the Sequential Importance Sampling (SIS) particle filter. This type suffers from a common problem called degeneracy phenomenon, where after a few iterations, all except one particle will have negligible weight. This phenomenon can be reduced by a good choice of the importance density and by adding a resampling step. The resampling step is added, after calculating the weights and estimating the posterior density function, in order to sample from the set $\{X_k^i, W_k^i\}$ and obtain $N_s$ new samples each with a weight of $1/N_s$. The low-weight particles will be eliminated accordingly. However, the resampling step requires calculating the sum of all weights in order to normalize them before resampling which would limit the opportunity to pipeline the whole algorithm steps [29]. Because of its importance for the particle filter operation, resampling techniques have been extensively researched and they were surveyed in [30]. Fig. 6 illustrates the main steps of the particle filter algorithm.

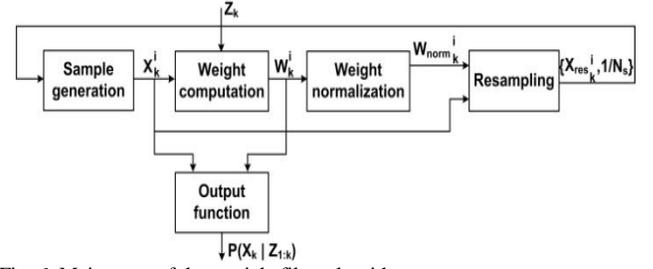

Fig. 6. Main steps of the particle filter algorithm.

Table II lists the filtering-based object trackers with HW implementations. The employed visual features are as simple as pixel intensity and more complex features like edges and color histograms. For the implementations of the pixel-intensity based particle filter, Li et al. [45] proposed a HW/SW co-design that implements the weight calculation step in SW on a NIOS-II CPU and all other steps in HW. A tournament selection approach was used for resampling to address the sample impoverishment problem. In addition, this allows avoiding the floating point operation of the weight calculation. The algorithm output is the highest-weight particle which simplifies the output stage. The proposed design achieves 50 fps and 13 fps for 32x32 and 64x64 mask size respectively with 32 particles on a Cyclone-2 FPGA.

Pixel intensity is also the employed visual feature in the complete HW implementation of the particle filter presented in [14]-[16]. Abd El-Halym et al. [14] presented three particle filter architectures: two-step architecture where the sample, weight and output steps are carried in parallel followed by a sequential resampling step. The second architecture is a parallel resampling step and the third architecture is a distributed particle filter with several processing elements and a central unit. In addition, the authors of this paper adopted a piecewise linear approximation of the exponential function to simplify the weight calculation step. Cho et al. [15] employed multiple features: Inter Frame Differencing (IFD) to detect objects and gray-level comparison to differentiate between two objects. The particle filter module is replicated to track two objects. Hong et al. [16] employed multiple motion models, constant velocity model for constant motion and current statistical model for maneuver motion, in order to track maneuvering targets while keeping fixed number of particles per model at all times. Independent Metropolis Hasting (IMH) was used in the resampling step where resampling can start processing the particles as they are generated and does not need to wait for the whole set of particles to be generated. Hence, all the particle filter steps can be pipelined which would reduce the execution time. After IMH generates a set of particles from each model, residual systematic resampling starts to determine how many particles each model contributes before the next sample step.

Other particle filter based trackers employ more complex visual features like edges [46], color histogram [47], [48]

and directional edges [49]. Alarcon et al. [46] proposed to extract edges using 3x3 Sobel filter and calculate the center of mass for the detected lane lines. Two lines only need to be stored accordingly. 25 fps was achieved with 12 particles on a cyclone FPGA. Wang et al. [47] proposed a color histogram based particle filter tracker. The authors adopted a YUV color histogram for the target and each of the particles with an L1 distance as a similarity measure between the histograms, and hence, particle weights were obtained. The authors reported that building the histograms and similarity measure take more than 90% of the computation time of the complete SW implementation that achieves 8 fps only. Therefore, the authors proposed a particle-level parallel computation of the histogram, where four particle histograms are computed in parallel, and pixel-level parallel processing, where the histogram spatial weights are calculated for two pixels in parallel and two histogram bins can be accumulated at the same time. The proposed design can estimate the object's position, size and angel. Second order dynamic model is employed in order to estimate the acceleration in addition to the velocity. The object's position is estimated by the average of the particle positions. However, the object's size is estimated by the best histogram similarity between nine particles of different sizes and the target. Same thing applies for the angle estimation where the authors proposed to divide the angle range into 32 entries and check the best similarity with the target. Synthesized with a UMC 90nm logic process, the proposed design can track objects of maximum size of 128x128 with 31.4 fps and 191 particles. Color histogram is also employed in a self-adaptive multi-threaded framework proposed by Happe et al. [48]. The proposed framework can adaptively switch between several HW/SW partitioning options during run time to react to the change of the input data and the performance requirements. The proposed platform consists of two PowerPC CPUs running SW threads at 300MHz and two reconfigurable HW slots running HW threads. Activating and deactivating cores can be done when tracking varying-size objects which would help in reducing the power consumption. The authors found that placing sampling or resampling steps in HW does not yield any improvement. Hence, these steps were not included in the partitioning decision while the observation step (e.g. histogram computation) and importance step (e.g. similarity measure) were included. The criterion to change the HW/SW partitioning is the performance (e.g. fps) where the system tries to keep the performance above a user-defined bound or within user-defined bounds. The authors adopted 100 particles divided into 10 chunks and reported a maximum performance of 40 fps on a Virtex-4 FPGA.

Zhao et al. [49] employed directional edges as the visual feature in order to be more robust against illumination variation and small variations in the object shape. The directional edge feature extraction consists of local and global feature extraction followed by an averaged principal-edge distribution step, which would produce 64-D feature vector as proposed in this paper. This vector is generated for the template and for each of the candidates. The tracking algorithms used is Multiple Candidate Regeneration (MCR) which is inherited from particle filter but with modifications and simplifications over the original particle filter algorithm. At initialization, the candidates are generated uniformly around the target location. Large weights are assigned for the candidates of large similarity with the target. Then, larger number of candidates is generated where large weights exist. The authors employed eight parallel processing blocks with Manhattan distance used as the similarity measure and a total of 64 candidates divided into eight groups. The eight candidates in each group are processed in parallel while the eight groups are processed serially. Consequently, the authors proposed two operation modes: high speed mode, where one group of candidates is only processed, and high accuracy mode, where all the candidate groups are processed. The proposed design can achieve up to 150 fps for 150x150 candidate size on a Stratix-3 FPGA.

TABLE II
FILTERING BASED TRACKER WITH HW IMPLEMENTATION

| Year/Tracker | Implementation | Visual feature | Image size | # of particles | Frame rate | HW Clock | MO | AD | Type |
|---|---|---|---|---|---|---|---|---|---|
| 2011 [45] | HW: Cyclone-2 SW: NIOS-2 | Pixel intensity | NR* | 32 | 50 fps | NR* | × | × | PF |
| 2012 [14] | HW: Virtex-5 | Pixel intensity | NR* | 64 | 270 k fps[1] | 36 MHz | × | × | PF |
| 2007 [15] | HW: Virtex-2 | Pixel intensity and IFD | 640x480 | 64 | 56.4 fps | NR* | ✓ | ✓ | PF |
| 2010 [16] | HW: Virtex-2 | Pixel intensity | NR* | 1024 | 50 k fps[1] | 60 MHz | × | × | PF |
| 2006 [46] | HW: Cyclone | Edges | 720x580 | 12 | 25 fps | 100 MHz | × | ✓ | PF |
| 2009 [47] | HW: UMC 90nm | Color histogram | 720x480 | 191 | 31.4 fps | 200 MHz | ✓ | × | PF |
| 2013 [48] | HW: Virtex-4 SW: 2 PowerPC | Color histogram | NR* | 100 | 40 fps | 100 MHz | × | × | PF |
| 2013 [49] | HW: Stratix-3 | Directional edges | 640x480 | 64 | 150 fps | 60 MHz | ✓ | × | MCR |
| 2015 [51] | HW: Zynq-7000 SW: Cortex-A9 | Pixel intensity | 1137x686 | NA** | 1.3 fps | 100 MHz | × | ✓ | KF |
| 2004 [50] | HW: EPXA10 SW: ARM9 | Edges | 1024x500 | NA** | 25 fps | NR* | ✓ | ✓ | KF |

[1] Estimated for algorithm only, does not include camera or memory access overhead
*NR: Not Reported in the original paper. **NA: Not Applicable.

For the Kalman filter based object trackers, Kaszubiak et al. [50] presented a 3D object tracker that takes images from two sources. Edge detection, correlation between two images and sub-pixel interpolation were done in HW in order to calculate the depth map while clustering of the depth map and Kalman filter for tracking were done in SW on an ARM9 CPU running at 166MHz. Liu et al. [51] also proposed a HW/SW co-design implementation of a Kalman filter based tracker where the Gaussian filter is only implemented in HW while adaptive background subtraction and Kalman filer are executed on two ARM Cortex-A9 CPUs running at 667MHz. The object's location and size are predicted.

To sum up the work on HW-based particle-filter trackers, it should be noted that the high frame rate reported in [14], [16] is the computation time for the particle filter only. Adding a camera interface or a memory interface would significantly affect this performance. Generally, the proposed designs in [14]-[16], which employ pixel intensity as the visual features, focus on optimizing the algorithm steps especially the resampling step because it typically prevents the possibility of a full pipelined implementation. However, if complex visual features like edges and histograms are employed in conjunction with particle filtering, then, the calculation of these features would become the performance bottleneck in the system, and hence, the focus should be on reducing the computation time of the feature extraction step that would significantly enhance the total system performance. In addition, it is evident that multi-model particle filter solutions achieve better tracking performance for real motion scenarios as presented in [16]. For more complex multi-model systems for instance, Kwon et al. [52] proposed a multi-model tracking system that can robustly work with several kinds of appearance and motion changes. This is done by employing several basic observation models and basic motion models with a basic tracker for each basic model as illustrated in Fig. 7. All basic trackers are integrated into a compound tracker in an interactive Markov Chain Monte Carlo framework. To address the increased computational complexity of the multi-model systems while achieving real time operation, we believe the research focus should be on the HW implementations of such systems.

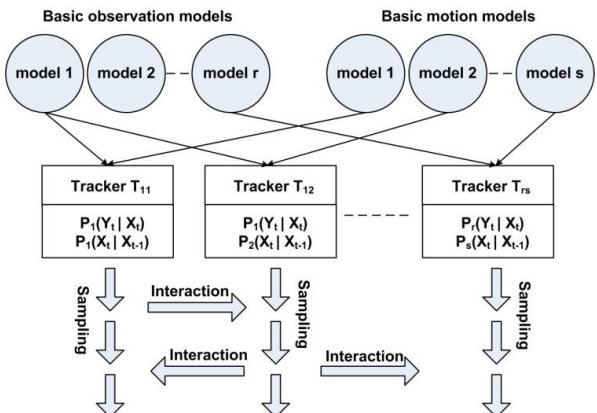

Fig. 7. The process of multi-modal tracker decomposition [52]

## C. Feature matching based tracker

The main idea of this category is to extract visual features from the current frame and attempt to correspond them with those extracted from the successive frames, and hence, the targets' positions can be estimated. A feature selection step typically exists in order to choose useful features only for further processing which would save the required computation time. The correspondence step is done in the feature space using several kinds of measures like absolute difference, Manhattan distance, Euclidean distance and Bhattacharyya measure.

Table III lists the feature matching based trackers with HW implementations. Muhlbauer et al. [53] and Ishii et al. [54] employed Harris corner detection technique to extract corner features from the images. Both papers used a 3x3 filter to obtain the image derivatives, $I_x$ and $I_y$. Muhlbauer et al. [53] employed non-maximum suppression over a window of 5x5 pixels as a feature selection technique. The window position is declared as a feature point if the center pixel is the maximum compared to all other pixels within the window. The Feature detection and selection were done in HW with six pipeline stages while the feature tracking based on normalized convolution was done in SW on a MicroBlaze CPU. Although the HW part can achieve 162 fps on a Virtex-2 FPGA and for 640x480 image resolution, the tracking part running in SW is the bottleneck in this system lowering the performance to 3 fps.

Ishii et al. [54], on the other hand, employed a threshold technique over a window of 8x8 as a feature selection technique. The maximum corner strength within the window is searched and if it is larger than a threshold value, this window will be considered as a candidate feature point. In addition, if the distance between the feature points is less than a pre-determined distance value, the feature with less corner strength will be discarded. In order to simplify the matching step, the authors utilized the high frame rate feature and assumed a small object displacement, and hence, they reduced the search region. The feature extraction was done in HW as it is a pixel-level operation while all other window-level operations were done in SW on a PC. The authors reported a high frame rate of 1000 fps for 1024x1024 video where the camera sends 16 pixels in parallel to the proposed design implemented on a Virtex-2 FPGA.

Color histogram based feature matching was presented in [55]-[57]. All have an RGB to HSV color conversion and generated histograms in the HSV color space. Cho et al. [55] proposed an adaptive color histogram based scheme where the appearance model of the target is blended with the appearance model of the most likely location, and hence, a new appearance model for the target is generated. For each pixel in the image, a histogram is generated for the Hue, Saturation and Value components from a window of 15x15 pixels centered on that selected pixel. All regions are compared with the reference model using Bhattacharyya similarity measure where the three similarities are blended to obtain a unified similarity. The authors employed 15 line buffers and a window buffer in order to generate and compare the histograms in a parallel and a pipeline structure, and hence, the required memory is saved. The proposed design can achieve 81 fps for VGA resolution on a Virtex-4 FPGA. Tomioka et al. [56] used Histogram of Oriented Gradient (HOG) in addition to the HSV histograms in order to obtain a close performance to the Scale Invariant Feature Transform (SIFT) features but with simpler HW implementation. The image is divided into 8x8 blocks where 32-bin HSV and 32-bin HOG histograms are generated for each block. Each block in the current frame is assigned exclusively to a block in the previous frame, background block or a new created block. This is done by solving a linear assignment problem using a savings-regret approximation instead of an optimal solution in order to simplify the HW implementation. In addition, sum of absolute difference was used instead of Bhattacharyya measure to get rid of the square root calculation in the Bhattacharyya computation. Using savings-regret approximation in addition to sum of absolute difference as a similarity measure cause a performance degradation of 2.38% on average. The authors proposed a 2D Processing Element (PE) array based on a Single Instruction Multiple Data (SIMD) architecture where each PE, which corresponds to a single block, is connected to four neighboring PEs. The authors proposed a Synchronous Shift Data Transfer scheme that allows each PE to communicate its data to any other PE. Moreover, object labeling process was performed using Connected Component Labeling (CCL) after solving the linear assignment problem. The proposed design can achieve 270 fps for 96x72 video resolution on a Virtex-3 FPGA while the complete SW implementation achieves 3 fps only. Gu et al. [57] divided the image into 8x8 cells as well to extract the histograms. However, the authors utilized the additivity feature of the color histogram calculation in order to obtain an object-level histogram. After generating the histogram of the Hue value of each cell, cell-based labeling process starts to label each cell based on the zeroth order moment feature. The histograms of the cells with the same label are summed to form an object-level histogram. Consequently, the matching of the object-level histogram is performed based on the Bhattacharyya similarity measure. The authors employed 16 eight-input histogram circuits as the camera sends eight pixels in parallel. The histogram and moment feature calculation were done in HW while the other steps were done in SW on a PC. The authors reported a high frame rate of 2000 fps for a 512x512 video resolution using the IDP express described before.

Yomaoka et al. [58] proposed a segmentation technique based on region-growing image-scan that exploits the large access bandwidth feature in the embedded memoires. This type of memories allows reading the results of segmenting one row of the image in one clock cycle. Simple features, position, size, area and color, are then extracted from the segmented regions and compared with those extracted from the preceding frame using Manhattan distance. A motion vector is also calculated to estimate the objects' locations in the next frame and reduce the search region. The proposed design can track up to 230 objects at 30 fps for 80x60 video resolution on a Stratix FPGA.

Hwang et al. [59] employed edges as the visual features that are extracted by a 3x3 Sobel filter. Three line buffers were used accordingly. To describe one feature point, four gradient values were calculated for the feature pixel in addition to its eight direct neighboring pixels. The authors set the maximum number of features points to be 20 features within an object bounding box of 50x50. Sum of absolute difference was used to compare the feature points in the search region, which is 100x100 pixels, in the next frame. A gravity center of the best matched points is considered the new location of the object. A PC was just used for video capture, user interface and video display. The proposed system achieves 30 fps for VGA resolution on a Spartan FPGA as the bottleneck is the host-FPGA communication while the proposed accelerators alone can support more than 100 fps.

TABLE III
FEATURE MATCHING BASED TRACKER WITH HW IMPLEMENTATION

| Year/Tracker | Implementation | Visual feature | Matching approach | Image size | Frame rate | HW Clock | MO | AD |
|---|---|---|---|---|---|---|---|---|
| 2006 [53] | HW: Virtex-2 SW: MicroBlaze | Corners | Normalized convolution | 176x144 | 3 fps | 100 MHz | ✓ | ✓ |
| 2009 [54] | HW: Virtex-2 SW: PC, Xeon Du. core | Corners | Absolute difference | 1024x1024 | 1000 fps | NR[*] | ✓ | ✓ |
| 2007 [55] | HW: Virtex-4 | Color histogram | Bhattacharyya similarity | 640x480 | 81 fps | NR[*] | ✗ | ✗ |
| 2014 [56] | HW: Virtex-3 (HW: Virtex-5)[1] | Color histogram | Sum of absolute difference | 96x72 (320x240)[1] | 270 fps (100 fps)[1] | 48 MHz (48 MHz)[1] | ✓ | ✓ |
| 2016 [57] | HW: Virtex-4 & Spartan-3 SW: PC, Core i7-975 | Color histogram | Bhattacharyya similarity | 512x512 | 2000 fps | 151.2 MHz | ✓ | ✗ |
| 2006 [58] | HW: Stratix (HW: Stratix-2)[2] | Position, size, color and area | Manhattan distance | 80x60 (320x240)[2] | 30 fps (30 fps)[2] | 20 MHz (20 MHz)[2] | ✓ | ✓ |
| 2014 [59] | HW: Spartan SW: PC[3] | Edges | Sum of absolute difference | 640x480 | 30 fps | NR[*] | ✗ | ✗ |
| 2013 [60] | HW: Zynq-7020 SW: Cortex-A9 | SURF | Nearest neighbor-ratio | 640x480 | 30 fps | 200 MHz | ✗ | ✓ |
| 2016 [61] | HW: Spartan-3 SW: Embedded CPU[2] | SIFT | Euclidean distance | 128x128 | 30 fps | NR[*] | ✓ | ✗ |

[1, 2]Results shown between brackets correspond to each other
[3]Undefined platform specification
[*]NR: Not Reported in the original paper.

More complex features which are scale and rotation invariant like Speeded Up Robust Feature (SURF) and SIFT were implemented in the tracker systems of [60] and [61]. Do et al. [60] partitioned the SURF algorithm into several sub-IPs connected to a shared bus as shown in Fig. 8. A rapid data transfer among the sub-IPs is performed by a Direct Memory Access (DMA). The SURF feature detection and description were done in HW while the feature matching was done in SW on a Cortex-A9 CPU. The feature matching is based on a nearest neighbor ratio matching algorithm where two vectors are matched if the ratio between the distance of the first nearest neighbor and the distance of the second nearest neighbor is less than a threshold. This helps in getting rid of potential incorrect matches. The proposed design achieves 30 fps for VGA resolution and 200x200 maximum object size on a Zynq platform. Yasukawa et al. [61] proposed a mixed analog-digital architecture to implement a SIFT based tracker. The authors adopted a MOS-based resistive network to perform the Gaussian filtering step. This resistive network would allow instantaneous filtering for the whole image and would minimize the power consumption. The other SIFT steps were done in a digital HW while the Euclidean distance based feature matching was done in SW. With the implementation on a Spartan-3 FPGA, 30 fps for 128x128 video resolution was achieved.

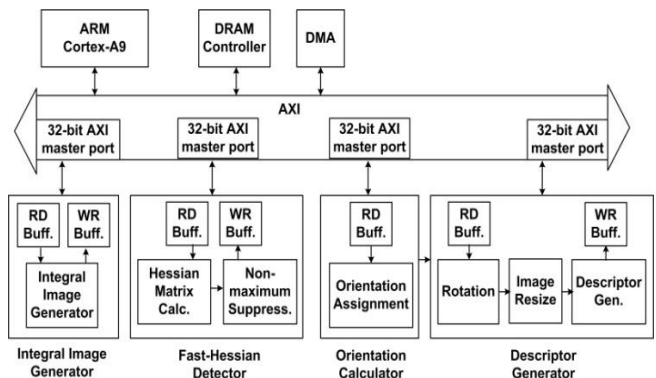

Fig. 8. Implementation of SURF-based tracker [60]

To sum up the work on HW-based feature-matching trackers, it is evident that the feature detection and selection steps are the computation bottleneck in the system, and hence, they are typically implemented in HW as it appears in all the published papers. The feature matching step, on the other hand, is much simpler and can be implemented in SW or HW based on the system requirements. High frame rate of 1000 fps and 2000 fps were reported in [54] and [57] respectively where the employed cameras output the pixels in parallel. Replicating the HW modules to process the parallel pixel streams allows achieving such high frame rate. In general, although several kinds of feature description techniques have been presented, no single one is robust and fast enough to deal with all tracking situations due to severe changes in appearance or motion [19]. Combining various complementary features with ensemble machine learning methods like boosting is a promising direction and the acceleration obtained from the HW implementation would be needed for such techniques.

On the other hand, learning visual features from deep neural networks is another approach to obtain more robust performance than that obtained by the hand-crafted features that are employed in all the surveyed papers. Deep neural networks are computation-intensive and would benefit from the HW acceleration.

### D. Optical flow based tracker

Optical flow is a dense field of displacement vectors that defines the translation of each pixel in a region [18]. In general, it assumes brightness constancy of corresponding pixels in consecutive frames. Various methods were proposed for the calculation of the optical flow vector. A popular method is the KLT tracker proposed by Lucas [62], Kanade [63] and Tomasi [64]. The main idea is to draw a window around a feature point and try to minimize the sum of square difference of the intensity of the window pixels in the current frame and those in the next frame. The sum of square difference is given by (10). Differentiating the sum of square difference and using a first order Taylor approximation give the optimum displacement in (11).

$$\varepsilon(d_x, d_y) = \sum_{x:-w_x}^{+w_x} \sum_{y:-w_y}^{+w_y} \left(I(x,y) - J(x+d_x, y+d_y)\right)^2 \quad (10)$$

$$[d_x \ d_y]^T = G^{-1}h \quad (11)$$

$$G = \sum_{x,y \in w} \begin{bmatrix} I_x^2 & I_x I_y \\ I_x I_y & I_y^2 \end{bmatrix} \ , \ h = \sum_{x,y \in w} \begin{bmatrix} I_x * (I-J) \\ I_y * (I-J) \end{bmatrix} \quad (12)$$

Where $I$, $J$, $I_x$ and $I_y$ are the first image, second image, derivative of image $I$ in the $x$ direction and $y$ direction respectively. $G$ is a 2x2 gradient matrix. The KLT scheme also proposed how to choose good features for tracking. This is done by choosing the feature points that have their eigenvalues of the $G$ matrix larger than a pre-defined threshold. This would indicate a visual corner. Calculating the two eigenvalues of $G$ is very similar to the Harris corner detection scheme. The basic optical flow steps are illustrated in Fig. 9. As first order Taylor approximation is used, this scheme would be valid only for small displacements. Hence, for more accurate results, an iterative KLT scheme has been typically used. In addition, in order to handle displacements larger than the window size, a pyramidal KLT tracker was proposed by Bouguet [65] where the basic KLT scheme is repeated for each scale with certain number of iterations.

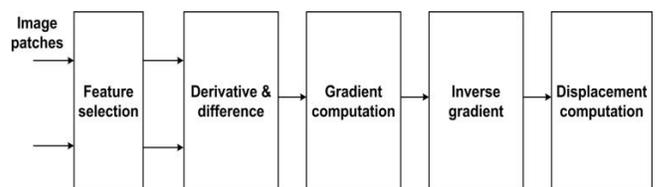

Fig. 9. Basic steps of the optical flow computation

Table IV lists the HW-based optical-flow trackers. Adaptive thresholding of the feature selection step was employed by Ghiasi et al. [66] and Jang et al. [67]. Adaptive thresholding would increase the system robustness against the temporal intensity variation. Ghiasi et al. [66] added a HW register for the threshold value which is loaded by a SW code running on a PowerPC CPU. The feature tracking step was also handled by the CPU while the preprocessing and feature selection steps were implemented in HW. This proposed HW/SW co-design system was actually implemented in several cameras in order to form a network embedded system. Real time operation was achieved on a Virtex FPGA using 3x3 window size and 150 features with 10% tolerance as a target of the total number of selected features. On the other hand, Jang et al. [67] employed histogram-based thresholding. The histogram is calculated for the current frame and a threshold is obtained and applied to the next frame which would facilitate a pipelined implementation. This is done under the assumption the intensity distribution is approximately stationary within the subsequent frame. The authors implemented a pyramidal KLT tracker where two off-chip memories were employed to store the image pyramids of the current and the previous frame. The maximum size of the memory is 4/3 frame size based on an infinite summation of a geometric series. A pipelined architecture was proposed where the features and the pyramids of the current frame are obtained while the tracking is taking place for the preceding two frames as illustrated in Fig. 10. 60 fps was achieved for 720x480 video resolution on a Virtex-5 FPGA.

Basic optical flow implementations were presented in Schlessman et al. [68] and Chai et al. [69]. Schlessman et al. [68] employed floating point arithmetic units and Symmetric Bipartite Table Method (SBTM) which performs second order Taylor approximation to approximate the reciprocal function. Using a Virtex-2 FPGA, 18 fps was achieved using a window size of 9x9. Chai et al. [69], on the other hand, depended on the high-frame rate capability of new cameras and assumed a small inter-frame displacement which would allow implementing the KLT algorithm in its basic form. The displacement is not more than 5 pixels. A non-maximum suppression was employed for the feature selection. The authors proposed a pipelined architecture with a memory divided into three sections. The gradients of the first and second frames are stored in two sections. When tracking between the first and second frame is performed, the gradient of the third frame is stored in the third memory section. The authors reported a performance of 182 fps for 1024x768 video resolution using a Virtex-5 FPGA.

To sum up the work on HW-based, optical-flow trackers, all the papers employed the KLT method with corners as the visual feature to implement optical flow based trackers. Basic implementations of the optical flow algorithm were proposed in [68], [69]. However, the proposed trackers in [66], [67] handle some issues of the optical flow algorithm like temporal intensity variation and small displacement which should result in a better tracker performance at the expense of more computations. Pipelined architectures were proposed by Jang et al. [67] and Chai et al. [69]. However, there would be one frame delay to output the result of the tracking which would not be acceptable for certain applications. In general, dealing with large displacement brings optical flow closer to feature matching techniques [34]. Several works listed in [34] have tried to combine the density and accuracy of the optical flow with the ability to capture large displacement of the feature matching technique. We believe focusing on the HW implementation of such techniques would overcome the main optical flow limitations while achieving better performance.

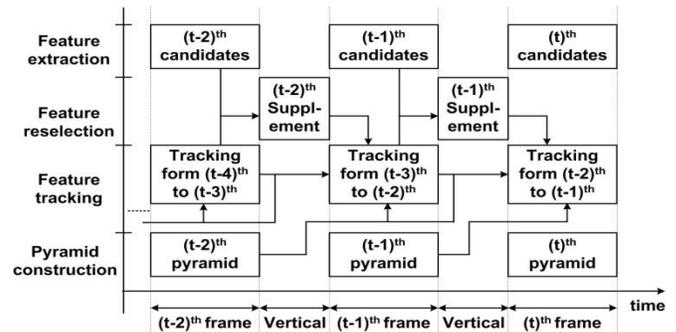

Fig. 10. Processing sequence of a pyramidal KLT tracker [67]

TABLE IV
OPTICAL FLOW BASED TRACKERS WITH HW IMPLEMENTATION

| Year/Tracker | Implementation | Visual feature | Image size | Frame rate | HW Clock | MO | AD |
|---|---|---|---|---|---|---|---|
| 2004/[66] | HW: Virtex SW: PowerPC | Corners | Width: 1280 | 30 fps | NR* | ✗ | ✗ |
| 2009/[67] | HW: Virtex-5 | Corners | 720x480 | 60 fps | NR* | ✓ | ✓ |
| 2006/[68] | HW: Virtex-2 | Corners | NR* | 18 fps | 67 MHz | ✗ | ✓ |
| 2011/[69] | HW: Virtex-5 | Corners | 1024x768 | 182 fps | 182 MHz | ✓ | ✓ |

*NR: Not Reported in the original paper.

*E. Template matching based tracker*

Template matching is the classical method in visual tracking. The main idea is to first represent the object by a template and then attempt to find the most similar region to this template in the next frames which would be considered the updated object locations. Several similarity measures have been used like Sum of Absolute Difference (SAD) (13), Standard Correlation (SC) (14), Normalized Correlation (NC) (15), and Normalized Cross Correlation (NCC) (16).

$$SAD(n,m) = \sum_{x,y \in T} |I(n+x, m+y) - T(x,y)| \quad (13)$$

$$SC(n,m) = \sum_{x,y \in T} I(n+x, m+y) T(x,y) \quad (14)$$

$$NC(n,m) = \frac{\sum_{x,y \in T} I(n+x, m+y)T(x,y)}{\sqrt{\sum_{x,y \in T} I^2(n+x, m+y)} \sqrt{\sum_{x,y \in T} T^2(x,y)}} \quad (15)$$

$$NCC(n,m) = \frac{\sum_{x,y \in T}[I(n+x, m+y) - \mu_I][T(x,y) - \mu_T]}{\sqrt{\sum_{x,y \in T}[I(n+x, m+y) - \mu_I]^2} \sqrt{\sum_{x,y \in T}[T(x,y) - \mu_T]^2}} \quad (16)$$

Where $T$, $I$, $\mu_T$ and $\mu_I$ are the template, current frame, mean value of the template and mean value of the image respectively. The minimum value of SAD or the maximum value of SC, NC and NCC indicates the best estimate of the new object location. SC is highly sensitive to illumination variation and the correlation value depends on the template size, while NC and NCC are less sensitive to illumination changes and the correlation value ranges from -1 to 1 at the expense of more computation complexity [70]. Furthermore, the basic template matching is a brute force method where the whole image is searched to obtain the best match to the template. This would lead to high computation cost [18]. Therefore, a search region close to the previous object location is typically determined for the algorithm to search in, and hence, mitigate the required computations.

Table V lists the HW-based template matching trackers. Dias et al. [71] employed SAD based template matching to test their proposed smart camera design. In the proposed design, a programmable control module which is executing a program code is connected to six PEs. The PEs are configurable window-based modules that can perform several 2D operations on the image patches based on the opcodes received from the control module. Fig. 11 illustrates the proposed configurable PE. The authors adopted a Stratix FPGA and reported a performance of 55.6 fps for template size 32x32 and search window 50x50. SAD was also employed by Adiono et al. [72] in a human detection application. The authors adopted Grayscaling and Binarization steps in order to simplify the SAD computation. SAD can be computed by just XOR operations on binary images. Hence, the authors designed a PE that can compute SAD of 100 pixels using XOR operations and binary tree adder. In addition, they designed a processing array (PA) that consists of 40 PEs in order to compute SAD for a template size of 40x100 pixels while achieving 162 fps on a Cyclone-2 FPGA.

Edges as visual feature and SC as template matching were employed by Samochin et al. [73] in their proposed design for a ball recognition and tracking. The authors adopted a simple color filtering to first detect a ROI for each potential object. Sobel filter was applied on each ROI to extract edges while shape filters were then applied to correlate the extracted edges with the pre-known object shape. This process is performed under the assumption that the object would not come out its ROI in the next frame. The authors employed a common input buffer and a shared delay line between the shape filters for an efficient HW implementation of the filters. The authors reported a 60 fps for 80x80 pixels template size and 100x100 pixels search window on a Vritex-4 FPGA.

Chen et al. [74] employed NCC for template matching in a product inspection application. The authors proposed to simplify the NCC computation and remove the square root computation. In addition, sub-sampling of four for the template and target image was performed in order to speed-up the NCC computation. The authors proposed the pipelined implementation shown in Fig. 12. Using a Vritex-6 FPGA, 30 fps was achieved for a template size ranging from 64x64 to 192x160 and a search window size ranging from 64x64 to 640x480. Rummele-Werner et al. [75] also used NCC-based template matching in addition to two other simple tracking algorithms, hotspot and centroid, in their proposed configurable tracking system. The authors utilized the dynamic and partial reconfiguration capability in the FPGA to exchange the tracking algorithm without stopping the tracking process. No much details is given about the NCC implementation while 25 fps was reported using a Virtex-4 FPGA.

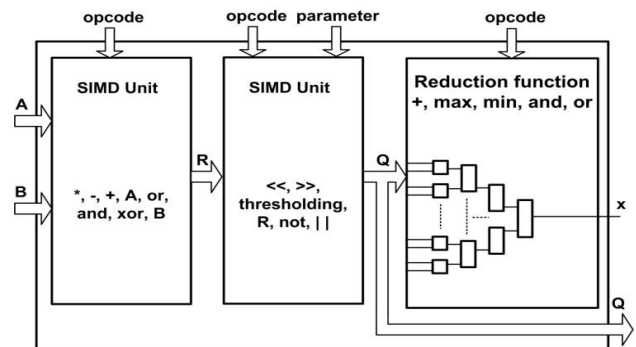
Fig. 11. Configurable window-based PE [71]

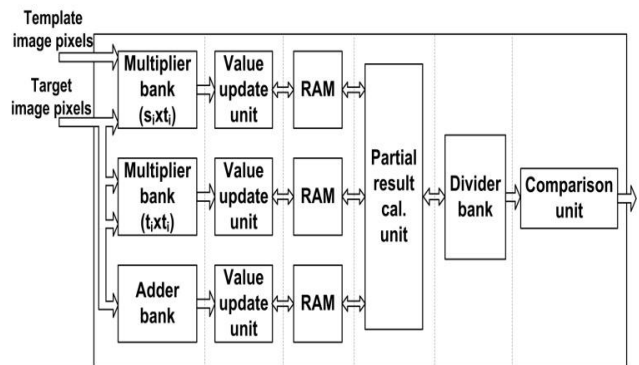
Fig. 12. Pipelined implementation of NCC computation [74]

TABLE V
TEMPLATE MATCHING BASED TRACKERS WITH HW IMPLEMENTATION

| Year/Tracker | Implementation | Visual feature | Matching approach | Image size | Template size | Frame rate | HW Clock | MO | AD |
|---|---|---|---|---|---|---|---|---|---|
| 2007 [71] | HW: Stratix | Pixel intensity | Sum of Absolute Difference | 2048x2048 | 32x32 | 55.6 fps | NR* | ✗ | ✗ |
| 2016 [72] | HW: Cyclone-2 | Pixel intensity | Sum of Absolute Difference | 640x480 | 40x100 | 162 fps | 50 MHz | ✗ | ✗ |
| 2010 [73] | HW: Virtex-4 SW: MicroBlaze | Edges | Standard correlation | 720x580 | 80x80 | 60 fps | 100 MHz | ✓ | ✓ |
| 2012 [74] | HW: Virtex-6 | Pixel intensity | Normalized cross correlation | 640x480 | 192x160 | 30 fps | 100 MHz | ✗ | ✗ |
| 2011 [75] | HW: Virtex-4 | Pixel intensity | Normalized cross correlation | 384x286 | NR* | 25 fps | 100 MHz | ✓ | ✗ |

*NR: Not Reported in the original paper

To sum up the work on the HW-based template-matching trackers, simple similarity measures like SAD was adopted in [71] and [72] while correlation-based similarity measure was adopted in [73]-[75]. Samochin et al. [73] extracted edges as the visual feature while all other trackers adopted pixel intensity. In general, the change in the object appearance causes a serious problem for the trackers based on template matching. Hence, updating the template would be mandatory for tracking objects when changing its appearance [1]. Ahmed et al. [70] proposed to update the template smoothly using first order IIR filter. In addition, correlation filters have achieved state-of-the-art performance on latest tracking datasets [1], [76]. The main idea is to model the appearance changes by an adaptive filter that is trained on image patches while the tracking step is carried out by convolution. Typically, this type of trackers exploits the frequency domain for the tracking and for the filter update in order to speed-up the computation. We believe that the use of correlation filters and/or using more than one appearance template for the object would be a promising direction for the future HW implementations of the template matching based trackers.

*F. Bio-inspired based tracker*

Bio-inspired techniques exploit biological phenomena for efficient tracking systems. To the best of our knowledge, particle Swarm Optimization (PSO) is the only technique that is implemented as HW-based trackers. PSO, which is inspired by birds searching for food, is a population-based stochastic process exploiting the phenomenon of swarm intelligence [21]. PSO optimization is done by maintaining a population of particles where each particle updates its position based on its best position in addition to the global best position among all particles. This process is repeated iteratively until a common converging point or the maximum iteration number is reached. The score of the particle position is tested with an objective function to create a fitness map for each particle. The best value in the particle fitness map is considered a local best and the best value out from all fitness maps of all particles is considered the global best. PSO is governed by simple update equations given by (17), (18) for the position and velocity respectively of particle $i$ at iteration $t$.

$$p_i(t) = v_i(t) + p_i(t-1) \quad (17)$$

$$v_i(t) = w\, v_i(t-1) + c_1 r_1 \big(best_i - p_i(t-1)\big) + c_2 r_2 \big(best_g - p_i(t-1)\big) \quad (18)$$

Where $best_i$ and $best_g$ are the best of particle $i$ and the global best position respectively. $w$, $c_1$ and $c_2$ are the inertia weight coefficient and acceleration coefficients respectively. $r_1$, $r_2$ are uniformly distributed random number in [0,1]. The flow chart of the PSO algorithm is shown in Fig. 13.

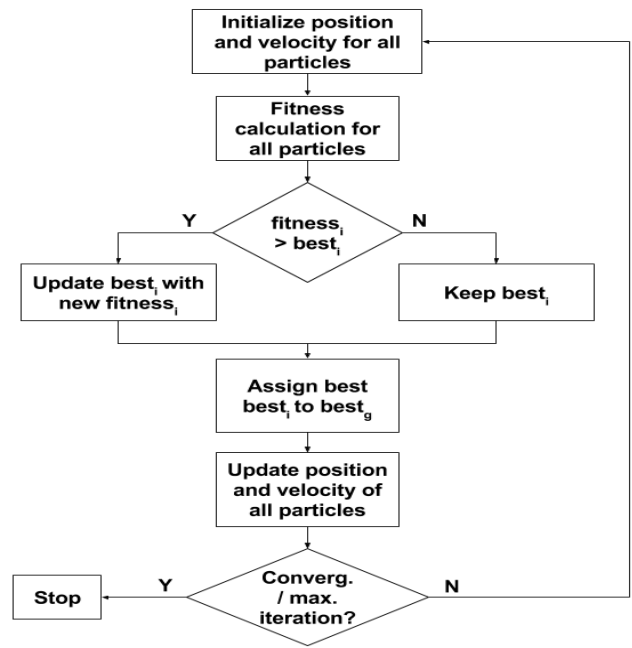

Fig. 13. Flow chart of the PSO algorithm

Table VI lists two PSO based trackers that feature HW implementation. Morsi et al. [77] employed Structural SIMilarity (SSIM) index as the fitness function which measures the similarity between the target and the particles based on luminance, contrast and structure. The authors designed three main components that operate sequentially for each particle: $best_i$ and $best_g$ calculation, velocity computation and position computation as illustrated in Fig.

14. A performance of 5.4 fps with 36 particles and 10 iterations was achieved using a Virtex-6 FPGA. Hsu et al. [78] proposed a hybrid particle filter and PSO system. The main idea is to use PSO in the beginning of the tracking in order to take advantage of the global optimization then switch to the particle filter. If during tacking the weights of the particles falls below a threshold, this would mean that the particle filter scheme fails and the system would switch to PSO till the particle weights become above another threshold. The authors proposed a HW/SW co-design where the weights are calculated in SW and the particle update is done in HW. The authors reported a performance of 12.4 fps and expected to reach 117 fps for a complete HW implementation.

It can be seen that low frame rates are achieved by both PSO-based designs because of the complex similarity function in [77] and the SW part in [78]. Generally, PSO is one of the contemporary approaches of the tracking algorithms as classified by [21]. There are several variants of PSO implementations and we believe it is a promising direction for researchers to focus on the HW implementations of such systems. Although PSO features simple update equations, the fitness function calculation would be the bottleneck in the system in addition to the number of particles and the storage required for each particle.

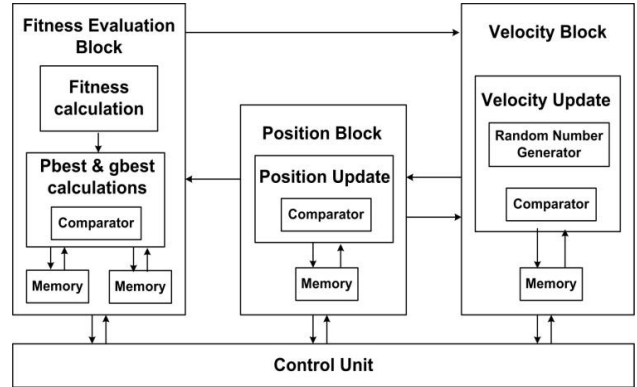

Fig. 14. Block diagram of a PSO implementation [77]

TABLE VI
PSO BASED TRACKERS WITH HW IMPLEMENTATION

| Year/Tracker | Implementation | Visual feature | # of particles | # of iterations | Frame rate | HW Clock | MO | AD |
|---|---|---|---|---|---|---|---|---|
| 2013/[77] | HW: Virtex-6 | Pixel intensity | 36 | 10 | 5.4 fps | 27 MHz | × | × |
| 2014/[78] | HW/SW | Pixel intensity | 32 | NR* | 12.4 fps | NR* | × | × |

*NR: Not Reported in the original paper

## IV. GENERAL DISCUSSION

We reviewed the published HW-based object trackers and classified them into six categories based on the tracking technique employed. In general, we do not aim in this survey to carry out one-to-one comparisons between the different HW-based trackers. It is evident that there are so many system parameters and trade-offs that obstruct reaching to a strong preference for the hardware implementation of the visual trackers. Currently, the diversity of tracking algorithms, lack of standard tracking performance measure, the diversity of the used HW platforms, and the lack of sufficient HW implementation details all combine to make direct comparison of HW trackers infeasible. Instead, we aim at describing the different HW architectures to give better insights for researchers and HW designers about the current status of the HW trackers and pave the road for future developments. We summarize our main findings from surveying the HW-based trackers in the following discussion.

It can be seen that extraction of complex visual features, like corners, edges, histograms, SURF and SIFT, is typically the performance bottleneck in the system, and hence, it has been commonly implemented in HW. The tracking algorithm, on the other hand, has been implemented in either HW or SW based on the performance target of the system. In general, pipelined implementations are adopted to speed up the computation at different system stages on the pixel-level, particle/object-level or frame-level. Pipelining is usually employed as long as the initial latency is affordable. In addition, parallel computations are also adopted through increasing the HW resources. Time-consuming functions like division, exponential and square root operations are commonly optimized or approximated for better performance.

We believe that system profiling is highly important before starting the efforts of the HW implementation. For example, optimizing the particle filter stages in a particle filter based system would not contribute much for the total system performance if complex features are adopted as they become the performance bottleneck consequently. Proper profiling would save the HW development time and enhance the overall system characteristics. In addition, fixed-point representation of the different system variables is commonly preferred in HW implementations. Therefore, proper adjustment of the fixed-point representation would save unnecessary extra logic and power while achieving the required performance.

The choice of the visual camera and camera interfacing is critical for the overall system operation and performance. The camera interfacing could be the main bottleneck in the system. Typically, the camera outputs the frame pixel values serially in a raster scan fashion with or without vertical and horizontal blanking periods. The serial output can be one pixel at a time or more than one pixel at a time for advanced high rate cameras. Depending on the algorithms employed and the HW architecture, the pixels can be processed once received from the camera which would save the buffering memory and achieve a low processing latency. Otherwise, line buffers and window buffers would be needed if there exists window-based operations in the employed algorithms. The number of the line buffers increases as the window

15height increases. A double buffering may be needed as well to store the current frame pixels while processing the previous frame stored pixels. If there exists blanking periods output from the camera, these periods can be utilized to handle some processing on the received pixels and reduce the required memory. Consequently, the choice and configuration of the visual camera can significantly enhance the total system characteristics.

It can be seen that almost all the trackers listed in this survey achieved at least real time operation, 30 fps. Furthermore, it was reported that some trackers achieved frame rates beyond 100 fps even reaching to 1000 fps [54] and 2000 fps [38], [57] using advanced high frame rate cameras. However, not many papers cover the real environment challenges as occlusions, changing appearance, changing motion models and illumination variations in addition to multi-object tracking. Therefore, we believe the future research should focus on handling these challenges utilizing the HW acceleration capabilities.

We have presented future directions recommendations of the HW implementation of the trackers in each category: annealing update of the kernel bandwidth [40] and fragment-based approach [42] in the mean-shift category, complex-visual features and multi-modal based particle filter in the filtering category, combination of complementary features with ensemble machine learning and usage of deep features in the feature matching category, fusion with feature matching for large displacements in the optical flow category, and finally, correlation filters and update of the template with more than one appearance model in the template matching category. We see as well that PSO category is a promising research direction because of its simplicity, and yet, not much work exists in this category.

Furthermore, classification-based trackers have gained popularity in recent years due to their efficacy in performance and simplicity of the classification task [21]. The idea is based on training a binary classifier to discriminate the target from the background. This category is called online tracking as well because the training data is not pre-known and has to be generated online during tacking. Moreover, deep neural networks have been used recently in the visual tracking domain and achieved competitive performance as well. They have been employed for object-background classifiers [79], similarity function learning [80] or for deep features extraction [81]. Classification-based techniques and deep neural networks have achieved state-of-the-art performance in the latest benchmark [1] and the VOT challenge [76]. To the best of our knowledge, there are no HW implementations published so far in this type of trackers which would be a good direction for researchers.

It is worth mentioning that the visual trackers compared in this survey are frame-based trackers where the visual camera sends the pixels to the processing module frame by frame. However, there is another type of the visual sensors which are non-frame based where the sensor outputs only the pixels that encounter intensity changes asynchronously. This type of sensors employs what is called asynchronous Address Event Representation (AER). The basic idea of AER is that the visual camera sends the address (e.g. event) of the changed-intensity pixels on a common bus. The frequency of an address appearance on the bus is proportional to the intensity change. These events can be then filtered and processed in the digital domain. This type of sensors has been adopted in object tracking applications like [82]-[84]. It is evident that this type of non-frame based trackers would require less processing and power consumption compared to the frame-based trackers.

Finally, we urge the researchers to carry out tracking performance measurements for their HW implementations. Emerging performance measures like OTB [1] and VOT [76], together with carefully selected test videos, started to show up recently and are gaining wide acceptance in the visual tracking field. Researchers are strongly invited to use these benchmarks to measure the performance of their HW-based trackers. We also urge the researchers to report enough HW/SW details to, at least, evaluate the time needed to calculate a MAC operation on their platform. This would enable meaningful comparisons across different HW implementations employing different tracking algorithms.

## V. CONCLUSION

In this paper, a survey on the HW implementations of the visual object trackers published in literature is presented. We focus on the complete HW and HW/SW co-design approaches and classify them into six categories based on the tracking algorithm employed. We attempt to point out the different issues in implementing the object trackers in embedded systems and how the published papers tried to solve them. We think this would give a better insight for researchers to design an efficient embedded tracker system. It can be concluded from our survey that real-time operation of a basic tracker can be achieved easily in each tracking category. However, employing complex visual features, multi-appearance models and multi-motion models would introduce performance challenges. In addition, more focus should be put towards overcoming the real environment challenges. We present recommendations for the future research direction in each tracking category and highlight the recent tracking approaches that lack HW implementations published in literature.